\documentclass[11pt,a4paper]{article}
\renewcommand{\title}[1]{\topsep=0pt\begin{center}\Large\bf#1\end{center}\vspace{12pt}} 
\renewcommand{\author}[1]{\topsep=0pt\begin{flushleft}\large\rm#1\end{flushleft}} 
\newcommand{\address}[1]{\topsep=0pt\begin{flushleft}\footnotesize\it#1\end{flushleft}\vspace{12pt}} 
\renewcommand{\date}[1]{\topsep=0pt\begin{center}\small#1\end{center}\vspace{12pt}} 

\usepackage{bbm}
\usepackage{fullpage}
\def\R{\mathbbm{R}}
\usepackage[T1]{fontenc}
\usepackage[latin1]{inputenc}

\begin{document} 
\title{On a  proof  of  the  collapse conjecture for  a  diagonal
  Bianchi type-IX vacuum space-time} 

\author{T. Charters}

\address{
 \'Area Departamental de Matem\'atica\\
 Instituto Superior de Engenharia de Lisboa\\ Rua Conselheiro 
 Em\'{\i}dio Navarro, 1, P-1949-014 Lisbon, Portugal\\[1em]
 Centro de Astronomia e Astrof\'{\i}sica da Universidade de Lisboa \\ 
Campo Grande, Ed. C8, 1149-016 Lisbon, Portugal
 \\email: {tca@cii.fc.ul.pt}}

\begin{abstract}
It is given a  simple proof  of  the  collapse conjecture for  a  diagonal
  Bianchi type-IX vacuum space-time. It is shown that the codimension of  
the infinity stable attractor, restricted to the anisotropy plane, is not zero,
thus proving that ``escape along a channel'' is impossible.
\end{abstract}

\section{Introduction}

We  present  a  simple  proof of  the  collapse
conjecture  for a  diagonal  Bianchi type-IX  vacuum space-time.   The
collapse conjecture  for a Bianchi type-IX space-time  has been proved
by Lin and Wald (1989-90) \cite{Lin + Wald: 1989,Lin + Wald: 1990}. We
shall  consider here  the simple  case of  a diagonal  Bianchi type-IX
universe  in   vacuum.   The  demonstration   presented  here  differs
significantly  from the  one given  in \cite{Lin  + Wald:  1989}.  The
collapse conjecture states that an initially expanding Bianchi type-IX
space-time can not expand for an infinite time. It should, in a finite
time, reach  a maximum  of expansion and  then begin to  contract. 
The proof given in \cite{Lin + Wald: 1989} that there
do not exist solutions which expand forever is
divided in two steps. In the first step it is proved that for any
solution which expands forever the dynamical trajectory in the
anisotropy plane must ``escape to infinity'' along one of the channels
of the Bianchi type-IX potential. The second step uses  detailed properties
of the equations of motion to show that such ``escape along a
channel'' is impossible.

The Bianchi type-IX model (dubbed mixmaster) has been intensively study over the
years. This model was investigated 
by Belinskii, Khalalatnikov and Lifshitz \cite{Belinsky:1970ew, Khalatnikov:1969eg} and Misner
\cite{Misner:1969hg}.  One of the most appealing properties is its chaotic
motion dynamics. The emergence of chaos helped the understanding of the
singularities in general relativity
\cite{Misner:1969hg,Cornish:1996yg,Motter:2000bg,ChaosGRNato}
and the debate is still  raging (see \cite{Motter:2000bg} for a clear insight). 

For historical reasons the approach to the study of the Bianchi model was to
rewrite Einstein equations to a Hamiltonian form (with a constrain). This was
intensively pursued by 
Misner \cite{Misner69,ChaosGRNato}, and several authors, for which a nice Hamiltonian system was
constructed, although this specific form of  Hamiltonian  was not used in the proof of the
collapse conjecture by Lin and Wald \cite{Lin + Wald: 1989,Lin + Wald: 1990}. 

Following Lin and Wald \cite{Lin + Wald: 1989,Lin + Wald: 1990}, we argue in
this paper in  favour of a alternative form of writing the equations of motion 
for the mixmaster universe, namely, the form of a canonical dissipative
 system \cite{Ebeling} (pg. 115, and references there in). Dissipative   systems
are common in general relativity 
\cite{Bogoyavlensky,Wainwright  +  Ellis} and in particular in Bianchi type
models. Under some conditions, for instance if one consider  expanding universes
solutions, one can show that
there exists functions that are monotonous decreasing along the solutions of the
Einstein equations of the
model under consideration due to  the monotonous behaviour of the determinant of
the metric (see, for instance, \cite{Charters:2001hi} for particular relevant examples). 

The basic idea in this work is to write the equations of
motion for the mixmaster model in terms of quantities relative to the overall
rate of expansion of the universe, or equivalent, to the Hubble
scalar \cite{Wainwright  +  Ellis}. This latter form
is a natural way of writing the equations of motion and clearly reveals the
dissipative nature of the 
Einstein equations. In fact one can look for the Bianchi type-IX equations as a
system of non-linear coupled and non-linear damped oscillators, where the
non-linear damping is proportional to the square root of the energy of an equivalent
mechanical system. This very specific form of dissipation is
 a clear signature  of its  dissipative nature. This is the source of the
 simplicity of the proof given in this work. By studying the dynamics
 relatively  to the overall rate of expansion of the universe one can 
 reduce the  problem of studying the ``escape along a channel'' to the  dynamics of a planar
 dynamical system in the anisotropy plane,  and show that
the codimension of  
the infinity stable attractor, restricted to the anisotropy plane, is not zero,
thus proving that ``escape along a channel'' is impossible.

Let us start by stating some general definitions.
 
The general Bianchi type-IX space-time has topology $\R\times { S}^3$,
with a simply transitive action of the isometry group ${SU}(2)$ on
 ${S}^3$ spatial slices.

It is described by the following metric \cite{Ryan + Shepley:1975},
\begin{eqnarray}
ds^2                   =                   -d\tau^2                  +
e^{2\alpha}\sum_{i,j=1}^3\left[e^{\bf\beta}\right]_{ij}d\sigma^i
d\sigma^j.
\end{eqnarray}
Here $\sigma^i$,  $i=1,2,3$, are  isometry invariant one-forms  on the
three-sphere  ${ S}^3$,  $\alpha$  is  a scalar,  and  $[\beta]$ is  a
traceless  $3\times  3$  matrix.   Both  $\alpha$  and  $[\beta]$  are
functions of the proper time $\tau$ only.

One should not forget that there is a particular relevant example of a vacuum
spatially-homogeneous  model which also has topology $\R\times { S}^3$,
with a simply transitive action of the isometry group ${SU}(2)$ on
 ${S}^3$  spatial slices, the Taub-NUT space-time. This model, as was shown by
Misner \cite{Misner69PRL}, does not possess a diagonal metric and although non-singular cannot be
extended (for details see, \cite{Ryan + Shepley:1975}, pg. 138). The  Taub-NUT
space-time represents a universe which evolves from an open universe, to closed
and an open universe again.  The dynamical behaviour of this model for the local rotation
symmetric case  can be studied by determining the 
explicit solution of the Einstein equations (same reference, pg. 139) which show
that the determinant of the metric is always non-null.

For Bianchi type-IX case the picture is not so simple. For  a  diagonal
space-time,  let $\beta_i$,  $i=1,2,3$,  denote  the 
diagonal  elements  of  the  matrix  $[\beta]$.   Only  two  of  these
quantities are independent since $\beta_1  + \beta_2 + \beta_3 = 0$ on
account of  the tracelessness of $[\beta]$. We  choose the independent
variables to  be\footnote{Note that Lin-Wald \cite{Lin  + Wald: 1989}
considered      the      variables      $\beta_+=\sqrt{6}b_1$      and
$\beta_-=\sqrt{6}b_2$.}
\begin{eqnarray}
b_1       &=&-\frac{1}{2\sqrt{6}}\beta_3,        \\       b_2       &=&
\frac{1}{6\sqrt{2}}\left(\beta_1-\beta_2\right),
\end{eqnarray}
which measure the departure from  isotropy.  In this case the Einstein
vacuum equations takes the form \cite{Ryan + Shepley:1975},
\begin{eqnarray}
\label{a}
3\dot  \alpha^2  =  \frac{1}{2}\left(\dot  b_1^2+\dot  b_2^2\right)  +
e^{-2\alpha}V(b_1,b_2),
\end{eqnarray}
\begin{eqnarray}
\label{b1b2}
\ddot   b_i  +   3\dot\alpha\dot   b_i  +   e^{-2\alpha}\frac{\partial
V}{\partial b_i} = 0, \qquad i = 1,2,
\end{eqnarray}
with
\begin{eqnarray}
\label{Vb1b2}
V(b_1,b_2)      &=&
-e^{-\sqrt{\frac{2}{3}}b_1}\cosh(\sqrt{2}b_2)\nonumber\\
\label{Vb1b2}
\quad +\frac{1}{4}e^{-4\sqrt{\frac{2}{3}}b_1} &&+
\frac{1}{2}e^{2\sqrt{\frac{2}{3}}b_1}\left[\cosh(2\sqrt{2}b_2)-1\right],
\end{eqnarray}
and
\begin{eqnarray}
\label{dda}
3\ddot\alpha +3\dot\alpha^2-\left(\dot b_1^2+\dot b_2^2\right) = 0.
\end{eqnarray} 
Note  that  only  two  of  the  three  sets  of  equations  (\ref{a}),
(\ref{b1b2}) and (\ref{dda}) are independent.

\section{Proof of the collapse conjecture}

Consider  the  new  time  variable $\displaystyle  \frac{dt}{d\tau}  =
e^\alpha$ monotonically  related to $\tau$  then, equations (\ref{a}),
(\ref{b1b2}) and (\ref{dda}), become,
\begin{eqnarray}
\label{a1}
3{\alpha'}^2 = \frac{1}{2}\left({b_1'}^2+{b_2'}^2\right) + V(b_1,b_2),
\end{eqnarray}
\begin{eqnarray}
\label{b1b2_1}
b_i'' + 2\alpha'b_i' + \frac{\partial  V}{\partial b_i} = 0 \qquad i =
1,2,
\end{eqnarray}
and
\begin{eqnarray}
\label{dda_1}
\alpha'' = -\frac{1}{3}\left({b_1'}^2+{b_2'}^2\right).
\end{eqnarray}

Equation (\ref{a1}) is a first  integral for the full six order system
defined by equations (\ref{b1b2_1}) and (\ref{dda_1}).

Let us define the total  energy function for the anisotropic variables
by
\begin{eqnarray}
\label{Energy}
E(b_1,b_2,  b_1',  b_2') =  \frac{1}{2}\left({b_1'}^2+{b_2'}^2\right)+
V(b_1,b_2),
\end{eqnarray}
and  note  that $E(0,0,0,0)<0$  because  $V(b_1,b_2)$  has a  negative
global   minimum  at   $(b_1,b_2)=(0,0)$. 
In   a  small   neighborhood  of   the  minimum
$(b_1,b_2)=(0,0)$ the potential (\ref{Vb1b2}) takes the form
\begin{eqnarray}
V(b_1,b_2) = -\frac{3}{4} + \left(b_1^2+b_2^2\right) + O_3(b_1,b_2).
\end{eqnarray} 

Let us see that the rate of expansion of the universe can not be positive for all
time. 
Consider  the
initial   conditions  $\alpha'(t_0)  >0$   (initially  expanding
universe), $\alpha(t_0)$ and $(b_1, b_2, b_1', b_2')(t_0)$
arbitrary. Let us consider that $\alpha' >0$ for all time in order to have a contradiction.
Because equations (\ref{b1b2_1}) are the equations of two non-linear damped
oscillators with damping factor $\alpha'>0$ it follows that the energy $E$, see (\ref{Energy}),
should be asymptotically negative,  because  $V(b_1,b_2)$  has a  negative
global   minimum  at   $(b_1,b_2)=(0,0)$, which is a contradiction because
$3E=\alpha'^2$. Then there exists  an instant  where $\alpha'  =0$ and 
thereafter $\alpha' < 0$.

Then  assume that $\alpha' < 0$ and consider the variables, which are the Hubble
normalised shear variables \cite{Wainwright  +  Ellis},
\begin{eqnarray}
  \label{eq:def_x}
  x_i=-\frac{b_i'}{\alpha'},\ i=1,2,
\end{eqnarray}
and the time variable change $d\xi =-\alpha' d\tau$, then equations (\ref{b1b2_1}) read
\begin{eqnarray}
\label{eq:x_xi_1}
  \frac{dx_i}{d\xi}&=& -2x_i+\frac{1}{3}x_i(x_1^2+x_2^2)-\frac{1}{V}\frac{\partial
    V}{\partial b_i}\left(3-\frac{1}{2}(x_1^2+x_2^2)\right),\\
\label{eq:x_xi_2}
  \frac{db_i}{d\xi}&=& x_i,\ i=1,2.
\end{eqnarray}

In order to study the escape along a channel it is useful to 
introduce the auxiliary variable $z_i=1/b_i$, $i=1,2$. We obtain
\begin{eqnarray}
  \frac{dz_i}{d\xi}=-z_i^2x_i.
\end{eqnarray}

Because the potential $V$ is invariant under rotations of $2\pi/3$ in the
anisotropic plane, see (\ref{Vb1b2}),  it suffices 
to study the escape in one of the channels. 
We chose the one defined by
$(b_1,b_2)=(b_1,0)$. 

Asymptotically, for large $b_1$ and $b_2\neq 0$  one has
\begin{eqnarray}
  \frac{1}{V}\frac{\partial V}{\partial b_i}\sim 2\sqrt{\frac{2}{3}}.
\end{eqnarray}
In this limit the $x_1$ dynamics decouples form the $x_2$
dynamics and the problem reduces to the study of a planar dynamical system of
the form
\begin{eqnarray}
\label{eq:ass.x_xi_1}
  \frac{dx_1}{d\xi}&=& 2\sqrt{6} -2x_1-\sqrt{\frac{2}{3}}x_1^2+\frac{1}{3}x_1^3,\\
\label{eq:ass.x_xi_2}
  \frac{dz_1}{d\xi}&=& -z_1^2x_1,
\end{eqnarray}
which shows that the infinity manifold $z_1=0$ is invariant
\footnote{For a similar argument applied to the study of scaling solutions in
  scalar fields models see \cite{Nunes:2000yc}.}. 
Equation
(\ref{eq:ass.x_xi_2}) shows that the stability of the invariant manifold
depends on the sign of $x_1$ in a small neighborhood of $x_1=0$, and thus the
codimension of  the stable manifold is not zero thus proving that escape along
this channel is impossible.

\section{Conclusions}

It was  given a  simple demonstration of  the collapse  conjecture for
vacuum  diagonal Bianchi type-IX  space-time by studying the dynamics 
 relatively  to the overall rate of expansion of the universe. Reducing the
 equations of motion in the anisotropic plane to a planar 
 dynamical system, it was shown that
 the infinity stable attractor does not have codimension  zero, and thus the 
 ``escape along a channel'' of the Bianchi-IX potential is impossible.

This  demonstration, for
this  case, is    simple  and concise and it provides  a new understanding
of this  classical cosmological model.

\end{document}